\documentclass{article}
\usepackage[a4paper, margin=1in]{geometry}
\usepackage{color}
\usepackage{booktabs}
\usepackage{comment}
\usepackage{amsmath}
\usepackage{tabularx}
\usepackage{hyperref}
\usepackage{authblk}
\usepackage{subcaption}
\usepackage{subcaption}
\usepackage{graphicx}
\hypersetup{
	bookmarks=true,         
	unicode=false,          
	pdftoolbar=true,        
	pdfmenubar=true,        
	pdffitwindow=false,     
	pdfstartview={FitH},    
	pdftitle={My title},    
	pdfauthor={Author},     
	pdfsubject={Subject},   
	pdfcreator={Creator},   
	pdfproducer={Producer}, 
	pdfkeywords={keywords}, 
	pdfnewwindow=true,      
	colorlinks=true,       
	linkcolor=blue,          
	citecolor=blue,        
	filecolor=magenta,      
	urlcolor=cyan           
}

\title{
Non-participant externalities reshape the evolution of altruistic punishment
}

\author[1]{Zhao Song}
\author[2]{Chen Shen}
\author[3]{Valerio Capraro}
\author[1]{The Anh Han}

\affil[1]{School of Computing, Engineering and Digital Technologies, Teesside University, United Kingdom}
\affil[2]{Faculty of Engineering Sciences, Kyushu University, Japan}
\affil[3]{Department of Psychology, University of Milan Biccoca, Italy}

\date{ }

\begin{document}

\maketitle

\begin{abstract}

While voluntary participation is a key mechanism that enables altruistic punishment to emerge, its explanatory power typically rests on the common assumption that non-participants have no impact on the public good. Yet, given the decentralized nature of voluntary participation, opting out does not necessarily preclude individuals from influencing the public good. Here, we revisit the role of voluntary participation by allowing non-participants to exert either positive or negative impacts on the public good. Using evolutionary analysis in a well-mixed finite population, we find that positive externalities from non-participants lower the synergy threshold required for altruistic punishment to dominate. In contrast, negative externalities raise this threshold, making altruistic punishment harder to sustain. Notably, when non-participants have positive impacts, altruistic punishment thrives only if non-participation is incentivized, whereas under negative impacts, it can persist even when non-participation is discouraged. Our findings reveal that efforts to promote altruistic punishment must account for the active role of non-participants, whose influence can make or break collective outcomes.

\end{abstract}

\section{Introduction}

The evolution of altruistic punishment, a key mechanism for sustaining cooperation, presents a long-standing theoretical puzzle~\cite{ sigmund2001reward,boyd2003evolution,raihani2012punishment}. While experimental evidences confirm that individuals will incur personal costs to punish non-cooperators, successfully sustaining cooperation in human societies, the evolutionary origin of this behaviour is difficult to explain~\cite{fehr2002altruistic, boehm1993egalitarian}. This puzzle stems from its inherent costs: punishers are not only vulnerable to exploitation by ``second-order free-riders"---cooperators who benefit from punishment without paying the cost~\cite{kaul2003providing,fehr2000cooperation,egas2008economics}---but also risk retaliation or antisocial punishment---
from those they sanction~\cite{denant2007punishment,janssen2008evolution,fehr2003detrimental}. To solve this paradox, researchers have proposed mechanisms, including group selection ~\cite{powers2013co,boyd1992punishment}, pool punishment ~\cite{sigmund2010social,szolnoki2011phase,perc2012sustainable}, and prior commitment ~\cite{han2016emergence,han2017evolution}, which typically rely on reciprocity or institutional rules to balance the costs and benefits of punishment.

Voluntary participation offers a powerful bottom-up approach for the evolution of punishment without reciprocity ~\cite{fowler2005altruistic,brandt2006punishing}. In classic models, this mechanism (i.e., non-participation strategy) provides an escape hatch, where players can opt out of the public good for a small, fixed payoff~\cite{hauert2002replicator,requejo2012stability}. This option destabilizes the dominance of defectors through a cycle dynamic where non-participation is more advantageous than mutual defection but less so than mutual cooperation. This environment prevents a total collapse to defection, allowing punishment to emerge through the neutral drift between punishers and non-punishing cooperators~\cite{hauert2007via}. However, this influential explanation rests on the critically narrow assumption that non-participants are passive actors, a simplification that overlooks their potential for more complex strategic behaviour.

In real-world social dilemmas, the decentralized nature of non-participation implies that an individual's freedom to opt out, with complex payoffs, can still have direct consequences for the public good. This active role stands in sharp contrast to the assumption of passive non-participation.
For instance, in public healthcare, an individual opting for private services may receive a net positive payoff (better care) as well as a negative one (higher fees)~\cite{meleddu2020public,hoel2003public}, while their departure simultaneously reduces overcrowding, a positive influence on the public system. Conversely, in a knowledge-sharing group, the departure of an influential member harms the group's collective expertise while offering the departing individual a potential gain (e.g., increase prestige) or loss (e.g., loss of collaborative opportunities)~\cite{tan2017let,das2013employee}. This real-world complexity motivates a critical re-examination of non-participation, raising key questions: Will this mechanism, when more broadly defined, remain effective in enabling the evolution of altruistic punishment? And under what conditions will it help or hinder it?

\begin{figure*}[htb]
    \centering
    \includegraphics[width=\linewidth]{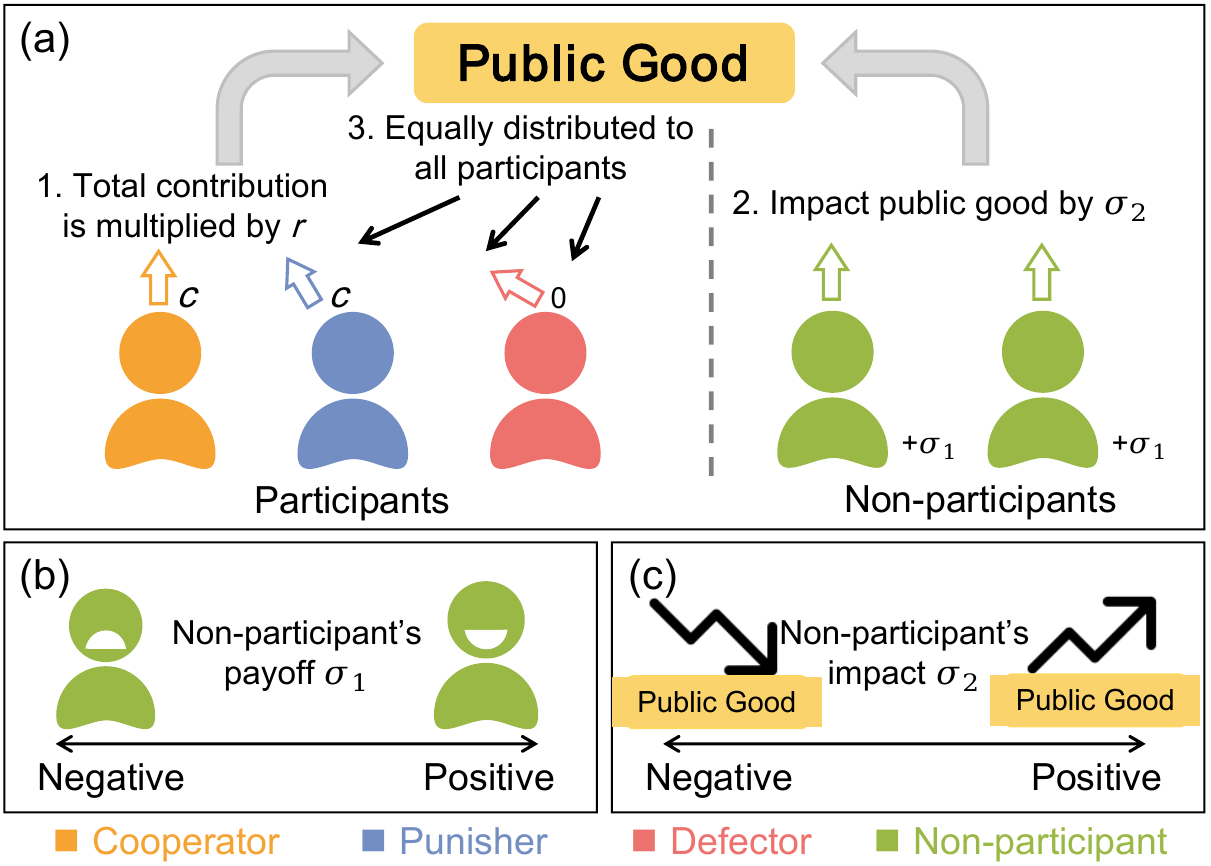}
    \caption{
    \textbf{Public goods game with voluntary participation.}
    (a) Each player has four options: cooperation, defection, punishment, and non-participation. Cooperators and punishers contribute an endowment $c$ to the public pool, while defectors contribute $0$. Non-participants opt out of the game with a payoff $\sigma_1$. All the contributions will be multiplied by the synergy factor $r$ as the public good. Each non-participant has an impact $\sigma_2$ on the public good. Then the public good is equally distributed among all participants, regardless of their initial contribution. (b) Non-participant's payoff $\sigma_1$ can be negative, positive, or zero. (c) Non-participant's impact on the public good $\sigma_2$ can be negative, positive, or zero.}
    \label{fig: model}
\end{figure*}

To answer these questions, we generalise non-participation in the public goods game by introducing a new parameter: the direct impact of non-participation on the public good ($\sigma_2$), reflecting a beneficial or harmful influence on the public good available to the remaining participants (see Figure \ref{fig: model}). Additionally, we extend the non-participation payoff ($\sigma_1$) beyond the non-negative range, representing the incentive or penalty for opting out. Alongside the non-participation, our model incorporates cooperation, defection, and altruistic punishment. In the game, cooperators and punishers contribute to the public pool, while defectors do not. The collective contribution in the public pool will be enhanced and then modified by the impact of all non-participants before being distributed equally among all participants.

Through evolutionary analysis in well-mixed finite populations, we find that the capacity of voluntary participation to support altruistic punishment depends critically on whether non-participants contribute positively or negatively to the public good. When opting out yields a positive payoff and non-participants exert a beneficial influence on the public good, altruistic punishment can dominate, and the range of synergy factors where it thrives expands compared to the traditional assumption that non-participants have no impact on the public good. In contrast, when non-participants undermine the public good, altruistic punishment becomes harder to sustain, and the viable synergy range narrows relative to the no-impact baseline. Notably, under positive impacts on the public good, the dominance of altruistic punishment requires positive 
incentives for non-participation, whereas under negative impacts, it can persist even when opting out is discouraged. These findings underscore the importance of accounting for the active role of non-participants when designing exit-based mechanisms to address the challenge of sustaining altruistic punishment.

\section{Model and Methods}

\subsection{Model}
In this study, we extend the assumption of non-participation within the framework of the public goods game (PGG), utilising one critical parameter: the direct impact $\sigma_2$, which represents the influence that non-participants have on the public good. Within the $N$-player PGG, each player has four distinct strategic options: cooperation ($C$), defection ($D$), punishment ($P$), and non-participation ($L$), as illustrated in Figure \ref{fig: model}(a). Cooperators and punishers contribute an endowment $c$ into the public pool, while defectors contribute nothing, and non-participants opt out of the game, receiving a payoff $\sigma_1$. First, all contributions are multiplied by the synergy factor $r$ as the public good. Each non-participant will impact the public good by $\sigma_2$. Then, the resulting public good is equally distributed among all players who have participated (i.e., excluding non-participants). 
Additionally, punishers incur a cost $\alpha$ to impose a loss $-\beta$ on defectors. Let $n_c$, $n_d$, $n_p$, and $n_l$ denote the number of players adopting each strategy, respectively, where $n_c+n_d+n_p+n_l=N$. Therefore, the payoffs for each strategy are given by:
\begin{equation}
    \begin{split}
        &\pi_C = \frac{(n_c+n_p)rc+n_l\sigma_2}{N-n_l}-c,\\
        &\pi_D = \frac{(n_c+n_p)rc+n_l\sigma_2}{N-n_l}-n_p\beta, \\
        &\pi_P = \frac{(n_c+n_p)rc+n_l\sigma_2}{N-n_l}-c-n_d\alpha,\\
        &\pi_L = \sigma_1.
    \end{split}
\end{equation}
Note that if only a single player chooses to participate, $N-n_l=1$, the PGG fails to be established, and the solitary participant also receives the non-participation payoff $\sigma_1$, as the other $n_L$ players.

As depicted in Figure \ref{fig: model}(b)-(c),  the non-participant's payoff, $-1\leq\sigma_1\leq 1$, represents the net outcome of opting out, where a positive value incentivizes non-participation and a negative value encourages participation; the non-participant's impact, $-1\leq\sigma_2\leq 1$, reflects their direct influence on the public good, where a positive value benefits the participants and a negative value induces loss. This framework moves beyond classic models that typically assume a fixed, positive payoff and zero impact for non-participants. For a clear comparison with previous models, e.g. in \cite{hauert2007via}, we set $c=1$, $\alpha=0.3$ and $\beta=1$, unless otherwise specified.

\subsection{Well-mixed finite population }
We analyze the evolutionary dynamics within a finite, well-mixed population of $M$ players. In each round of the game, a group of $N$ players is randomly sampled from the population to play the PGG. Each player in the population can adopt one of the four strategies $C$, $D$, $P$, and $L$. Since players are sampled from a well-mixed finite population, the probability of forming a group with a specific composition of strategies follows a multivariate hypergeometric distribution. 
For a simplified case with two strategies, $i$ and $j$, where there are $m_i$ players of strategy $i$ in the population, the probability of sampling a group with $n_i$ players of strategy $i$ and $N-n_i$ players of strategy $j$ is given by:
\begin{equation}
H(n_i,N,m_i,M)=\frac{\binom{m_i}{n_i}\binom{M-m_i}{N-n_i}}{\binom{M}{N}}.
\end{equation}
The expected payoff for a player is calculated by averaging over all possible compositions of the $N-1$ co-players they might interact with in a sampled game group (see section \ref{Appendix A}). For example, in a population with $x$ cooperators and $M-x$ defectors, the expected payoffs for a focal cooperator and a focal defector are:
\begin{equation}
\begin{split}
P_{CD} &=\sum_{n_x=0}^{N-1}H(n_x,N-1,x-1,M-1)\left(\frac{(n_x+1)r}{N}-1\right) \\
&=\frac{r}{N}\left(\frac{N-1}{M-1}(x-1)+1\right)-1, \\
P_{DC} &=\sum_{n_x=0}^{N-1}H(n_x,N-1,x,M-1)\left(\frac{n_xr}{N}\right) \\
&=\frac{r}{N}\frac{N-1}{M-1}x.
\end{split}
\end{equation}
The evolution of strategies is modeled using a Moran process with a pairwise social learning mechanism. In each time step, one player $i$ is randomly chosen for strategy revision, and another player $j$ is chosen as a potential role model. The probability that player $i$ adopts player $j$'s strategy is determined by their fitness difference according to the Fermi function~\cite{sigmund2010social}:
\begin{equation}
p_{i\rightarrow j}=\frac{1}{1+\mathrm{exp}[-s(P_{ji}-P_{ij})]},
\end{equation}
where $P_{ij}$ and $P_{ji}$ are the expected payoffs (fitness) of players $i$ and $j$, respectively. The parameter $s$ represents the intensity of selection, which controls how strongly imitation depends on fitness differences. For $s=0$, imitation is random (neutral drift), while for $s\rightarrow \infty$, players always imitate strategies with higher fitness.

To determine the long-term success of each strategy, we first calculate the fixation probability, $\rho_{ij}$, which is the probability that a single mutant of strategy $i$ will eventually take over a resident population of $M-1$ players of strategy $j$. The transition probabilities $T_{ij}^{\pm}$ for the number of $i$ players to increase or decrease by one are given by:
\begin{equation}
T_{ij}^{\pm}=\frac{M-m_i}{M}\frac{m_i}{M}\frac{1}{1+\mathrm{exp}[\pm s(P_{ji}-P_{ij})]}.
\end{equation}
The fixation probability $\rho_{ij}$ is then calculated as~\cite{traulsen2006stochastic}:
\begin{equation}
\rho_{ij}=\frac{1}{1+\sum_{k=1}^{M-1}\prod_{m_i=1}^{k}\frac{T_{ij}^- (m_i)}{T_{ij}^+ (m_i)}}.
\end{equation}
Assuming a small mutation limit, where any mutant either fixates or goes extinct before another mutation occurs \cite{nowak2004emergence,imhof2005evolutionary},   the fixation probabilities $\rho_{ij}$ define the transition probabilities of the Markov process between four different homogeneous states of the population. The transition matrix with $T_{ij,i\neq j}=\rho_{ij}/(q-1)$ and $T_{ii}=1-\sum_{j=1, j\neq i}^qT_{ij}$, where $q$ is the number of strategies. The normalised eigenvector associated with the eigenvalue 1 of the transposed transition matrix provides the stationary distribution described above, describing the relative time the population spends adopting each of the strategies.

\subsection{Risk dominance}
To analyze the short-term invasion dynamics between any two strategies, we use the concept of risk dominance. Strategy $i$ is risk dominant against strategy $j$ when it satisfies \cite{sigmund2010calculus}:
\begin{equation}
    \sum_{k=1}^{N}\pi_{ij}\leq \sum_{k=0}^{N-1}\pi_{ji},
\end{equation}
where $\pi_{ij}$ and $\pi_{ji}$ are the payoff of strategy $i$ and $j$, respectively, when the random sampling consists of $k$ player $i$ and $N-k$ player $j$. 

\section{Results}

\begin{figure*}[tb]
    \centering
    \includegraphics[width=\linewidth]{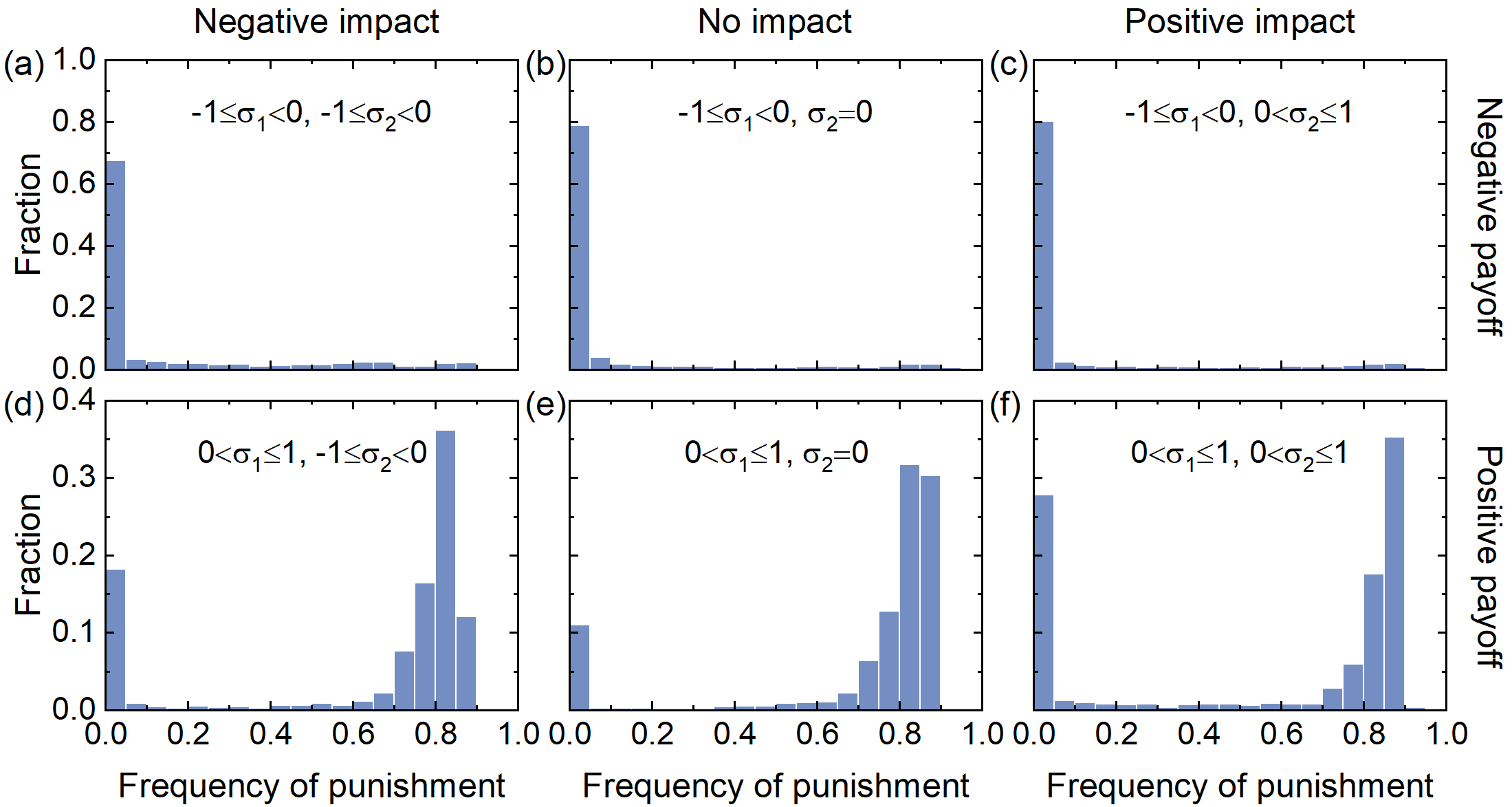}
    \caption{
    \textbf{Compared with the case without voluntary participation, the effectiveness of non-participation for promoting the emergence of punishment is limited.}
    Shown are the results of 10,000 numerical calculations for $s=1$. Parameters are randomly sampled from uniform distributions on the intervals $\alpha\in[0,1]$, $\beta \in [\alpha,5]$, $r\in[1,5]$, and (a) $\sigma_1\in[-1,0)$,  $\sigma_2\in[-1,0)$, (b) $\sigma_1\in[-1,0]$,  $\sigma_2=0$, (c) $\sigma_1\in[-1,0)$,  $\sigma_2\in(0,1]$,(d) $\sigma_1\in(0,1]$,  $\sigma_2\in[-1,0)$, (e) $\sigma_1\in(0,1]$,  $\sigma_2=0$, (f) $\sigma_1\in(0,1]$,  $\sigma_2\in(0,1]$.
    }
    \label{fig: robustness}
\end{figure*}

\begin{figure*}[htb]
    \centering
    \includegraphics[width=\linewidth]{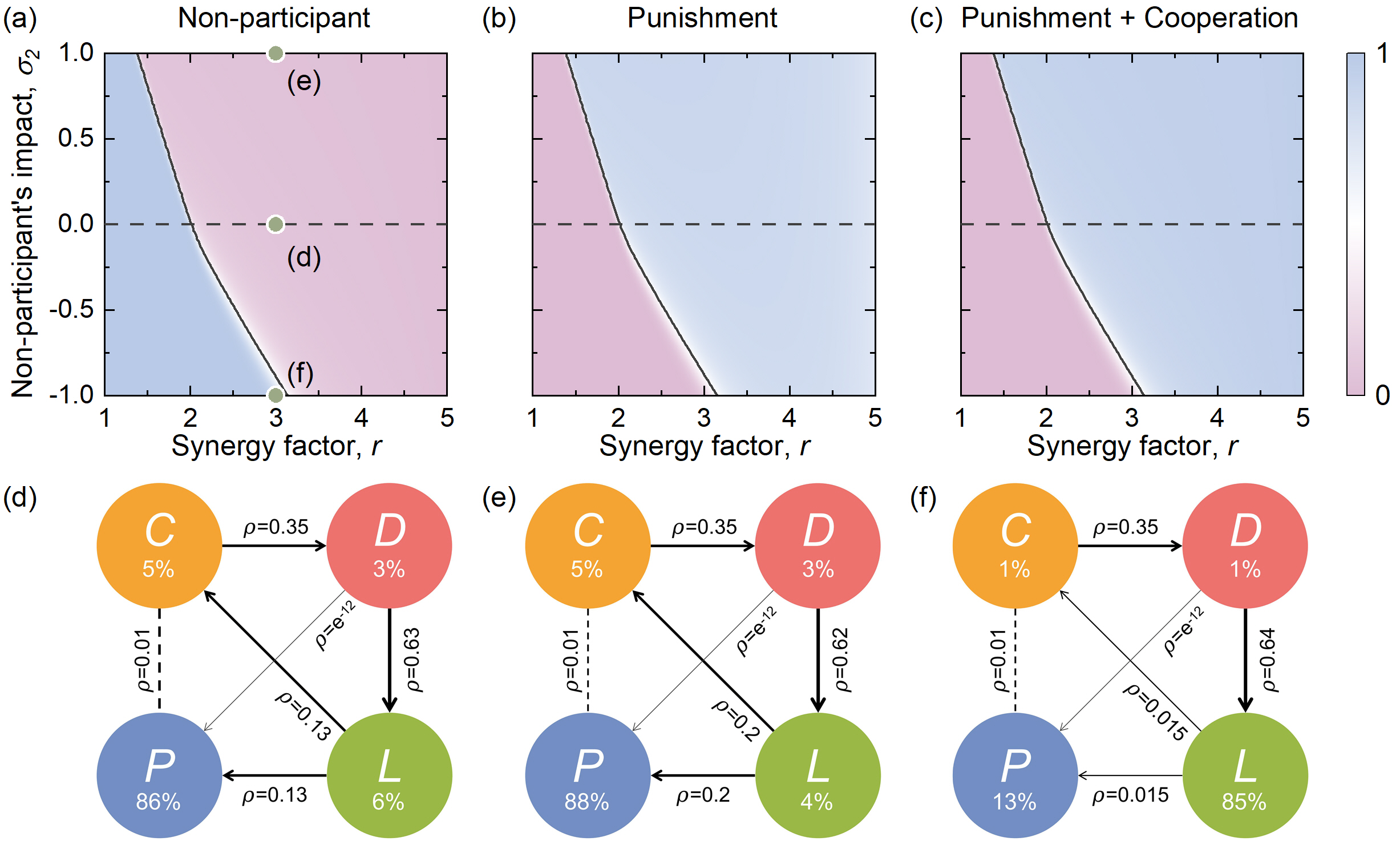}
    \caption{
    \textbf{Compared with the case of a negligible non-participation impact, a positive impact lowers the threshold of the synergy factor ($r$) for punishment survival, while a negative impact increases it. }
    Panels (a)-(c) show the frequency of non-participation, punishment, and the amount of cooperation and punishment as a function of the synergy factor and the non-participant's impact, respectively.
    The dashed line represents the case of non-participation having no impact on the public good. The solid line marks the threshold beyond which the indicated strategy dominates, i.e., with a frequency of at least 50$\%$.
    Panels (d)-(f) show the stationary distribution and the transition probabilities for the selected games indicated by the green circles in (a).
    Black arrows show the stronger transitions within a pair of strategies,  dashed arrows show neutral transitions, and $\rho$ denotes the transition probability.
    Parameters are set as $s=1$, $\sigma_1=1$, $r=3$, (d) $\sigma_2=0$, (e) $\sigma_2=1$, and (f) $\sigma_2=-1$.}
    \label{fig:fig1}
\end{figure*}

\begin{figure*}[htb]
    \centering
\includegraphics[width=\linewidth]{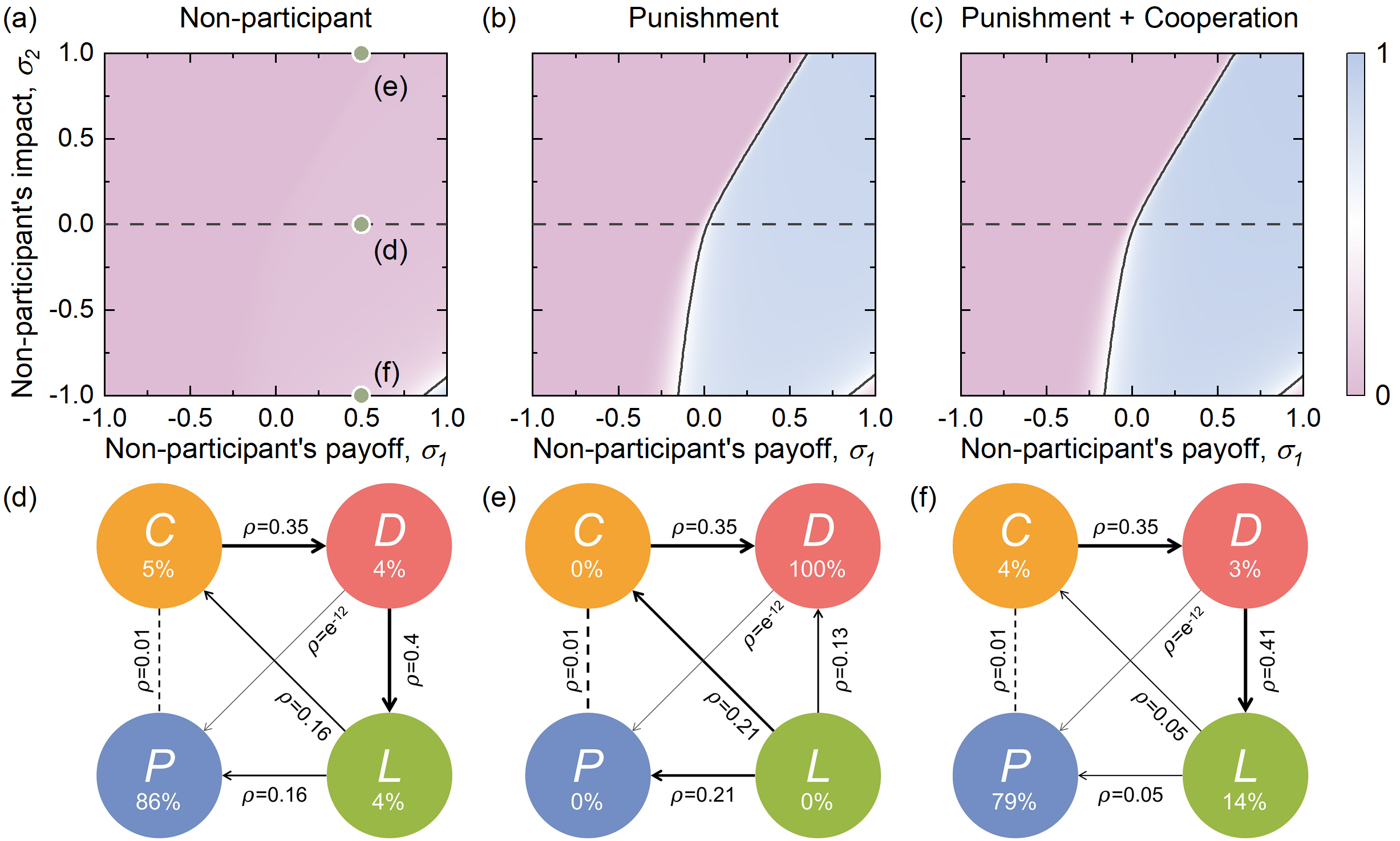}
    \caption{
    \textbf{A positive impact from non-participants requires a higher payoff incentive for punishment to dominate, while a negative impact allows punishment to persist even when non-participation is disincentivized.}
    Panels (a)-(c) show the frequency of non-participation, punishment, and the amount of cooperation and punishment as a function of the non-participant's payoff and the non-participant's impact, respectively.
    The dashed line represents the non-participation without the impact on the public good. The solid line marks the threshold beyond which the strategy dominates, meaning its stationary distribution probability is more than 50$\%$.
    Panels (d)-(f) show the stationary distribution and the transition probability for the selected points in (a), respectively.
    The black arrow shows the stronger transition between the two strategies, the dashed arrow shows the neutral transition, and $\rho$ denotes the transition probability.
    Parameters are set as $s=1$, $\sigma_1=0.5$, $r=3$, (d) $\sigma_2=0$, (e) $\sigma_2=1$, and (f) $\sigma_2=-1$. 
    }
    \label{fig:fig2}
\end{figure*}

\subsection*{Non-participation has a conditional impact on the evolution of altruistic punishment}
We first present our general finding that compared with the case without the opting out option, the effectiveness of non-participation in promoting the emergence of punishment is highly conditional.
This finding is robust, as evidenced by our examination of the stationary distribution of punishment averaged over 10,000 randomly sampled parameter sets, as shown in Figure \ref{fig: robustness} (see also supporting information Figure \ref{figA1}).

The non-participant's payoff, $\sigma_1$, plays the most decisive role ~\cite{han2016emergence,rand2011evolution}. When this payoff is negative (the top row in Figure \ref{fig: robustness}), punishment is strongly suppressed, with its frequency remaining near zero in the vast majority of parameter settings, regardless of the non-participant's impact. Conversely, a positive payoff (the bottom row in Figure \ref{fig: robustness}) is a necessary condition for punishment to evolve. Within this positive payoff regime, however, the non-participant's direct impact plays a significant and intricate role, creating polarized outcomes where punishment can either flourish or fail completely. In essence, a positive payoff is essential for punishment to emerge, but the nature of the non-participant's impact then determines the balance between its risk of extinction and its ultimate success. 

To better understand this general finding, below we provide a detailed analysis of how the positive or negative impact of non-participation can significantly influence evolutionary dynamics, contrasting with the no-impact setting typically considered in previous studies \cite{sigmund2010social,hauert2007via}.

\subsection*{Impact of non-participants alters the synergy threshold for altruistic punishment to dominate}
First, compared to the case without impact, the direct impact of non-participants significantly alters the synergy factor required for punishment to dominate the population, with the positive impact lowering the threshold and the negative impact raising it. Compared to the scenario with no impact ($\sigma_2=0$, dashed line in Figure \ref{fig:fig1}(a-c)), where punishment prevails when $r\geq2$, a positive impact considerably extends this range for the dominance of punishment, lowering the required synergy factor to around $r>1.4$. Conversely, a negative impact restricts these conditions, requiring a larger synergy factor to $r>3$. Detailed analysis of risk dominance is shown in section \ref{Appendix B}.
Additionally, punishment is the priority dominant cooperative behaviour, rather than cooperation (the dashed line, Figure \ref{fig:fig1}(b-c)). 

The underlying evolutionary dynamics reveal that the stability of a cyclic dominance among $C$, $D$, and $L$ is the key mechanism for the dominance of punishment. In the case without impact ($\sigma_2=0$, Figure \ref{fig:fig1}(d)), punishment is strongly favoured, comprising 86$\%$ of the stationary distribution. When a positive impact is introduced ($\sigma_2=1$, Figure \ref{fig:fig1}(e)), the dominance of punishment is slightly enhanced to 88$\%$, while non-participation is reduced. In contrast, a negative impact completely reverses this outcome ($\sigma_2=-1$, Figure \ref{fig:fig1}(f)). Non-participation becomes a highly stable state, making up 85$\%$ of the stationary distribution. Although the transition from non-participation to punishment is weak ($\rho$= 0.015), the reverse transition towards non-participation is even weaker. This slightly maintains the cyclic dynamics among $C$, $D$, and $L$, leading to a state of non-participation with a minimal level of punishment surviving ($13\%$).

\subsection*{Impact of non-participants influences their required incentives to achieve the dominance of altruistic punishment}
Second, compared to the no-impact setting, to achieve the dominance of punishment, a positive impact requires higher incentives for non-participants, while a negative impact does not, even when non-participation is disincentivised.
In the scenarios with no impact ($\sigma_2=0$, dashed line in Figure \ref{fig:fig2}), punishment dominates over a wide range, prevailing as long as the non-participant's payoff $\sigma_1$ is larger than 0. However, a positive impact severely restricts these conditions by raising the $\sigma_1$ threshold, meaning punishment can only dominate if the incentive to opt out is stronger. For instance, at a high positive impact ($\sigma_2=1$), punishment cannot dominate unless its payoff is larger than 0.6.
Conversely, a negative impact creates a different trade-off. It allows punishment to be maintained at a slightly broader range of $\sigma_1$ values compared to the baseline, but it also causes punishment to fail if the payoff $\sigma_1$ becomes strongly negative (around $\sigma_1<-0.9$). 

This is because a positive impact can destroy the cyclic dynamics that induce punishment. With a moderate payoff ($\sigma_1=0.5$) and no impact  ($\sigma_2=0$, Figure\ref{fig:fig2}(d)), punishment thrives (86\%) because a strong cycle reliable suppresses defection. However, when a strong positive impact is introduced ($\sigma_2=1$, Figure\ref{fig:fig2}(e)), punishment collapses completely and defection take over (100$\%$). The positive impact fundamentally breaks the enforcement cycle: the ability of non-participants to invade and replace defectors is severely disrupted, where the transition from defection toward non-participation is reversed. Without this crucial mechanism, the population succumbs to defection. In addition, when a negative impact is considered ($\sigma_2=-1$, Figure\ref{fig:fig2}(f)), though the transition from non-participation toward cooperation and punishment is weakened ($\rho$ drops from 0.16 to 0.05), the cyclic dynamics exist, and therefore, punishment prevails.
These results indicate that, though the positive impact on public good benefits participants, it potentially destroys the cyclic dynamics where non-participation loses the advantages when competing with defection, therefore, the extinction of punishment. Furthermore, the negative impact decreases public good, while the advantage of non-participation towards defection is the key to sustaining the cyclic dominance. 
Our findings further underscore the limitation of voluntary participation in sustaining punishment, considering the impact of non-participants on public good.

\subsection*{Robustness across strong selection intensity}
\begin{figure*}[tb]
    \centering
    \includegraphics[width=\textwidth]{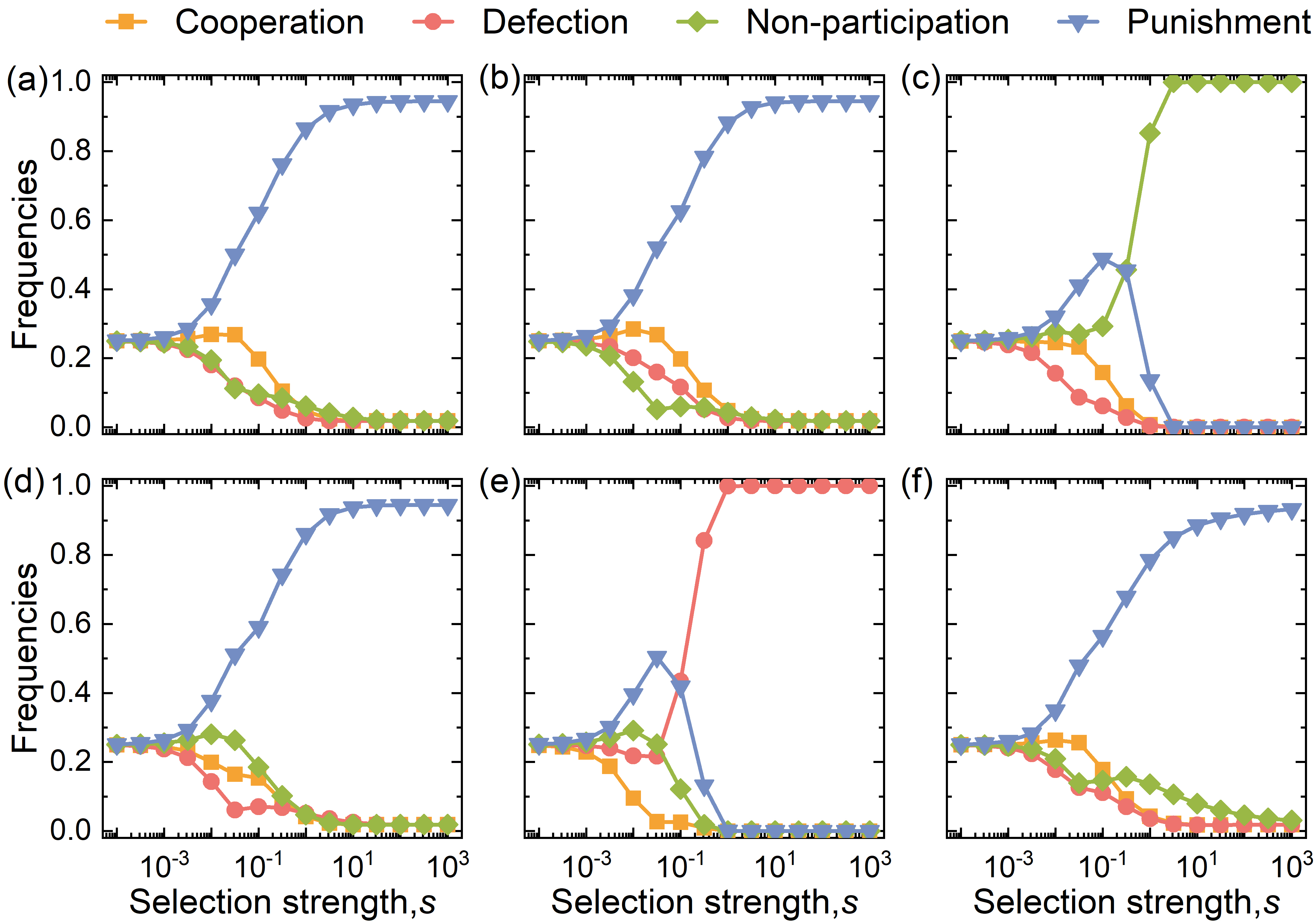}
    \caption{ 
    \textbf{The robustness across strong selection scenarios. }
    Stationary distribution as a function of selection strength $s$.
    Parameters are set as $r=3$, (a) $\sigma_1=1$ and $\sigma_2=0$, (b) $\sigma_1=1$ and $\sigma_2=1$, (c) $\sigma_1=1$ and $\sigma_2=-1$, (d) $\sigma_1=0.5$ and $\sigma_2=0$, (e) $\sigma_1=0.5$ and $\sigma_2=1$, and (f) $\sigma_1=0.5$ and $\sigma_2=-1$. }
    \label{fig:fig3}
\end{figure*}

Our findings are robust across the strong selection scenarios. 
In detail, when the non-participant payoff is high ($\sigma_1=1$, the top row in Figure \ref{fig:fig3}), a negative impact (($\sigma_2=-1$, Figure \ref{fig:fig3}(c))) leads to the dominance of non-participation under strong selection, while neutral or positive impacts lead to the dominance of punishment (Figure\ref{fig:fig3}(a-b)). Similarly, for a moderate payoff ($\sigma_1=0.5$, the bottom row in Figure \ref{fig:fig3}), a positive impact (($\sigma_2=1$, Figure \ref{fig:fig3}(e))) uniquely leads to the dominance of defection under strong selection, while neutral and negative impacts sustain the dominance of non-participation, where the later decreases the frequency of punishment slightly (Figure \ref{fig:fig3}(d)(f)).
Further results across selection strength for different synergy factors $r$ are shown in Figure \ref{fig:fig3_S1} and \ref{fig:fig3_S2}.
In all these cases, weak selection leads to the coexistence of four strategies (around $s<1$). These findings align with our previous conclusion and further highlight the careful implementation of voluntary participation as a promoter of punishment.

\section{Discussion}

In this work, we have re-examined the role of voluntary participation in sustaining altruistic punishment in the one-shot public goods game. To this end, we have extended the consequences of non-participation by a crucial parameter: its direct impact, which can be beneficial or harmful to the public good. Additionally, we have expanded the payoff for non-participants to a broader range, which acts as either an incentive or a penalty for opting out. Our results reveal that the effectiveness of non-participation is highly conditional. On the one hand, compared to the case without impact, the positive impact lowers the threshold of synergy factors for punishment to thrive, while the negative impact raises it. On the other hand, to achieve the dominance of punishment, the positive impact demands a higher incentive for non-participants, while the negative impact does not, even when non-participation is disincentivised. These findings emphasise the limitations of viewing voluntary participation as a universal promoter of punishment.

Our findings build upon previous studies by showing that the effect of voluntary participation in promoting punishment is highly dependent on the assumption of passive non-participants~\cite{hauert2007via,sigmund2010social}.
We demonstrate that the cyclic dynamics inducing punishment are fragile and can be disrupted by the impact of non-participation, whether positive or negative, on the public good. For example, our results indicate that a positive impact from non-participants, while seemingly beneficial, can break the cycle dynamics by reversing the transition from defection to non-participation, leading to the prevalence of defection. In addition, a negative impact can weaken the ability of punishers and cooperators to suppress non-participants, resulting in a decrease in punishment.
This result aligns with a broader re-evaluation of voluntary participation in sustaining cooperation, highlighting its limitation in solving social dilemmas given the non-negligible impact on the public good ~\cite{khatun2025optional}.
In other words, our findings underscore the necessity of carefully considering voluntary participation as a universal solution for social dilemmas from the evolutionary theoretical perspective.

The practical implications of these findings can be exemplified in real-world scenarios like open-source software development. The departure of an influential lead developer, for instance, represents a negative impact, which our model shows can trap a project in a non-productive state of low participation. Conversely, if non-essential members leave, it could be seen as a positive impact by reducing coordination costs; however, our results warn this could paradoxically weaken enforcement against low-quality contributions. Furthermore, high exit costs (e.g., losing access to a critical platform), corresponding to a negative payoff, would severely suppress the sanctioning behaviours needed to maintain project standards. To ensure voluntary participation benefits the public good, it is therefore crucial to manage the impacts of non-participation and structure incentives appropriately.

Finally, we acknowledge the key simplifications in our model—the consideration of a single, homogeneous type of non-participant within a well-mixed population. By demonstrating that even this simplified, singular type of non-participant can fundamentally alter and disrupt cooperative outcomes, our findings show the critical need for further investigation into more complex scenarios. First, it would be valuable to investigate hybrid populations containing both adaptive human players and simple, committed bots. While "loner bots" (who always opt out) in the optional prisoner's dilemma have been shown to have no impact on cooperation in well-mixed populations ~\cite{sharma2023small}, it remains an open question whether our non-participant bots would facilitate altruistic punishment in this context.
Second, the role of population structure should be explored. Future work could move from the current well-mixed setting to pairwise networks, where loner bots are known to facilitate cooperation~\cite{sharma2023small} and cooperative behaviours can cascade through human social interactions~\cite{fowler2010cooperative}. An even more advanced step would be to consider higher-order networks, which more faithfully represent group interactions and can uniquely promote prosocial behaviours in ways that simple networks cannot~\cite {majhi2022dynamics}. Lastly, our framework could be applied to other games involving moral behaviours to explore how the stability of norms depends on the externalities created by non-participants~\cite{capraro2021mathematical,kumar2020evolution}. Such work would further refine our understanding of how voluntary participation can be effectively managed to support, rather than undermine, cooperation.

\section*{Acknowledgments}
We acknowledge the support provided by EPSRC (grant EP/Y00857X/1) to Z.S. and T.A.H., and JSPS KAKENHI (Grant no. JP 23H03499) to C.\,S..


\section*{Competing interest} Authors declare that they have no conflict of interest.

\section*{Data availability}
No datasets were generated or analysed during the current study. The code to support the findings of this
study is available at \url{https://osf.io/tprkw/?view_only=1950c421dbca4bacb1734708c1cbc865}. 

\bibliographystyle{unsrt}
\bibliography{mybib}

\newpage
\setcounter{figure}{0}
\setcounter{equation}{0}
\renewcommand*{\thefigure}{A\arabic{figure}}
\renewcommand*{\theequation}{A\arabic{equation}}

\section{Appendix A} \label{Appendix A}
For the population, we denote by $x$, $y$, $z$, and $w$ the numbers of cooperators, defectors, non-participants, and punishers, respectively, where $x+y+z+w=M$. Among the selected players, if only one player participates in the game, the player acts as a non-participant, which happens with probability $\tbinom{z}{N-1}/\tbinom{M-1}{N-1}$. 
Otherwise, the expected payoffs for pairwise encounters are: 
\begin{equation}
   \begin{split}
        &P_{CD} =\frac{r}{N}\left(\frac{N-1}{M-1}(x-1)+1\right)-1, \\
        &P_{DC}=\frac{r}{N}\frac{N-1}{M-1}x,\\
       &P_{CP}=P_{PC}=r-1, \\
       &P_{CL}=P_{PL}=\left(1-\frac{\tbinom{z}{N-1}}{\tbinom{M-1}{N-1}}\right)\!\left(\frac{N\sigma_2}{\tbinom{M-1}{N-1}}\sum_{n_i=1}^{N-1}\frac{\tbinom{M-z-1}{n_i}\tbinom{z}{N-1-n_i}}{n_i+1}-\sigma_2+r-1\right)+\frac{\tbinom{z}{N-1}}{\tbinom{M-1}{N-1}} \sigma_1,\\
       &P_{LC}=P_{LD}=P_{LP}=\sigma_1, \\
       &P_{DL}=\left(1-\frac{\tbinom{z}{N-1}}{\tbinom{M-1}{N-1}}\right)\!\left(\frac{N\sigma_2}{\tbinom{M-1}{N-1}}\sum_{n_i=1}^{N-1}\frac{\tbinom{M-z-1}{n_i}\tbinom{z}{N-1-n_i}}{n_i+1}-\sigma_2\right)+\frac{\tbinom{z}{N-1}}{\tbinom{M-1}{N-1}} \sigma_1,    \\
       &P_{DP}=\frac{(N-1)(r-N\beta)}{N(M-1)}w,\\
       &P_{PD}=\frac{(N-1)(r+N\alpha)}{N(M-1)}(w-1)+\frac{r}{N}-1-\alpha(N-1).
   \end{split}
   \label{eq: payoff}
\end{equation}

\section{Appendix B} \label{Appendix B}
As there exists a neutral transition between cooperation and punishment, then we derive the condition for punishment to be risk-dominant against non-participation and defection, respectively.
\begin{itemize}
    \item Punishment is risk-dominant against defection when
    \begin{equation}
        \sum_{k=1}^{N}[\frac{rck}{N}-c-(N-k)\alpha] \geq \sum_{k=0}^{N-1} (\frac{rck}{N}-k\beta),
    \end{equation}
    which is simplified as 
    \begin{equation}
        \beta -\alpha \geq \frac{2c(N-r)}{N(N-1)}.
    \end{equation}
    \item Punishment is risk-dominant against non-participation when
    \begin{equation}
        \sigma_1 + \sum_{k=2}^{N}(\frac{rck+(N-k)\sigma_2}{k}-c) \geq \sum_{k=0}^{N-1}\sigma_1,
    \end{equation}
    which is simplified as 
    \begin{equation}
       rc-c+\frac{H_GN-2N+1}{N-1}\sigma_2\geq \sigma_1,
       \label{P-N}
    \end{equation}
    where $H_G = \sum_{k=1}^{N} \frac{1}{k}$.
\end{itemize}

Besides, we derive the condition for non-participation to be risk-dominant against cooperation and defection, respectively.
\begin{itemize}
    \item Non-participation is risk-dominant against cooperation when 
    \begin{equation}
        \sum_{k=1}^{N}\sigma_1 \geq \sigma_1 + \sum_{k=0}^{N-2}(\frac{rck+k\sigma_2}{N-k}-c),
    \end{equation}
    which is simplified as
    \begin{equation}
         \sigma_1 \geq \frac{N H_G-2N+1}{N-1}(rc+ \sigma_2) - c.
    \end{equation}
    \item Non-participation is risk-dominant against defection when 
    \begin{equation}
        \sum_{k=1}^{N}\sigma_1 \geq \sigma_1 + \sum_{k=0}^{N-2}\frac{k\sigma_2}{N-k},
    \end{equation}
    which is simplified as 
    \begin{equation}
        \sigma_1 \geq  \frac{N H_G - 2N + 1}{N-1}\sigma_2.
    \end{equation}
\end{itemize}
Accordingly, as the parameters set in Figure \ref{fig:fig1} and Figure \ref{fig:fig2}, the condition where punishment is risk-dominant against other strategies can be obtained.
\newpage

\newpage
\section{Appendix C}
\begin{figure}[htb]
    \centering
    \includegraphics[width=0.9\linewidth]{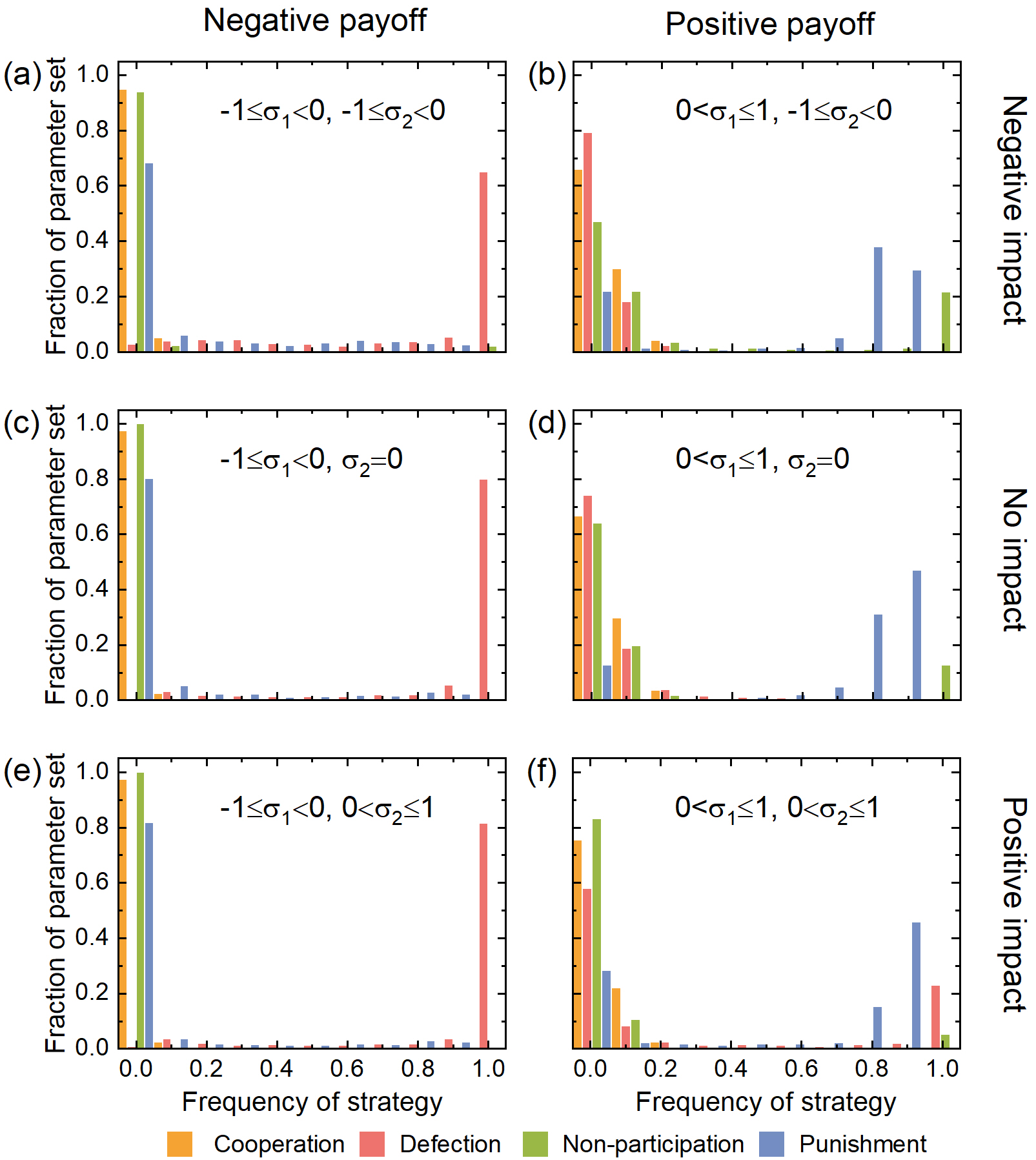}
    \caption{
    \textbf{The effectiveness of non-participation for promoting the emergence of punishment is limited.}
    Shown are the results of 10,000 numerical calculations for $s=1$.
    Parameters are set as $\alpha\in[0,1]$, $\beta \in [\alpha,5]$, $r\in[1,5]$, $\sigma_1\in[-1,1]$, and $\sigma_2\in[-1,1]$. }
    \label{figA1}
\end{figure}




\begin{figure}[htb]
    \centering
    \includegraphics[width=\linewidth]{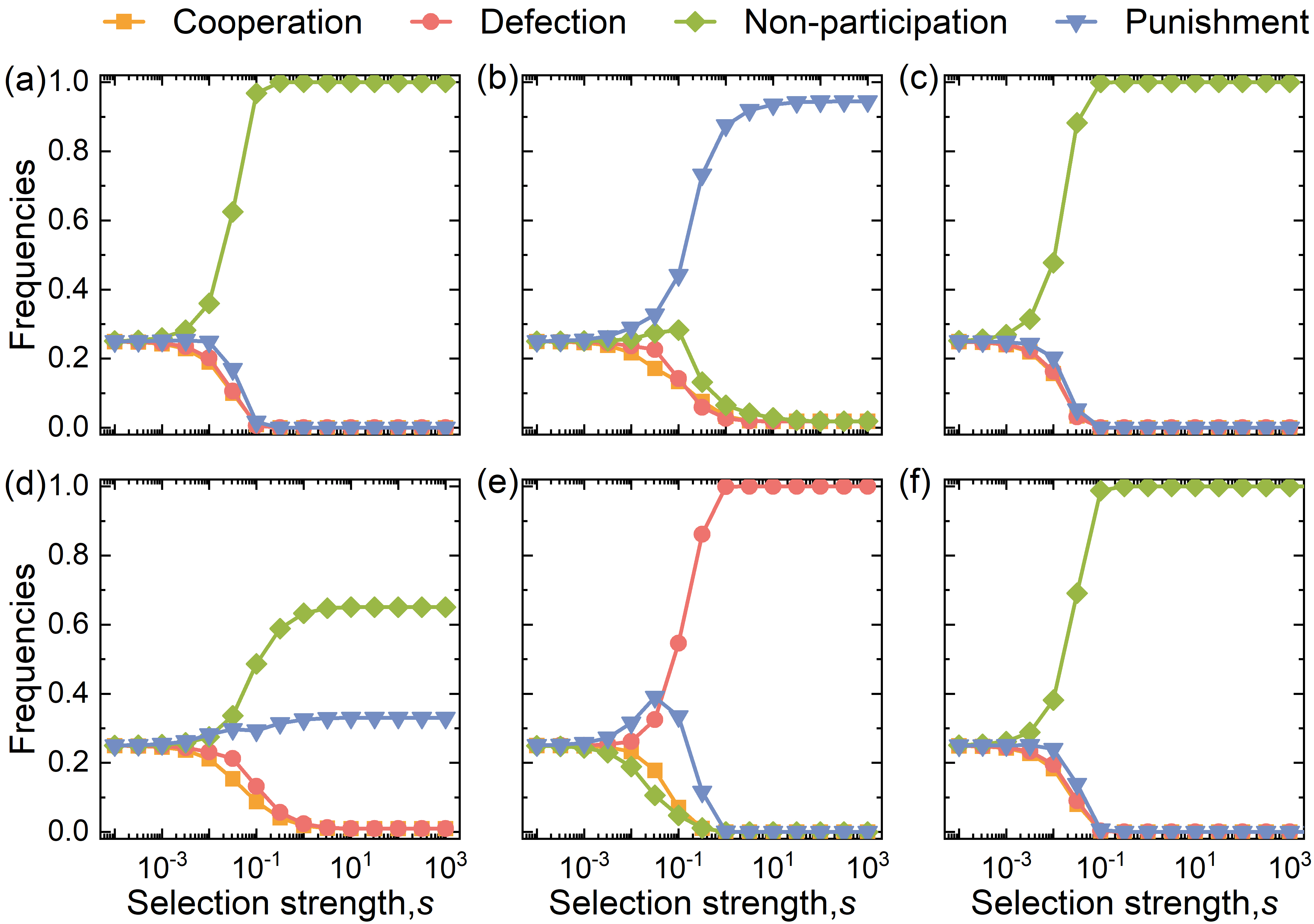}
    \caption{
    \textbf{The robustness across strong selection scenarios. }
    Shows are the stationary distribution as a function of selection strength $s$.
    Parameters are set as $r=1.5$, (a) $\sigma_1=1$ and $\sigma_2=0$, (b) $\sigma_1=1$ and $\sigma_2=1$, (c) $\sigma_1=1$ and $\sigma_2=-1$, (d) $\sigma_1=0.5$ and $\sigma_2=0$, (e) $\sigma_1=0.5$ and $\sigma_2=1$, and (f) $\sigma_1=0.5$ and $\sigma_2=-1$. }
    \label{fig:fig3_S1}
\end{figure}

\begin{figure}[htb]
    \centering
    \includegraphics[width=\linewidth]{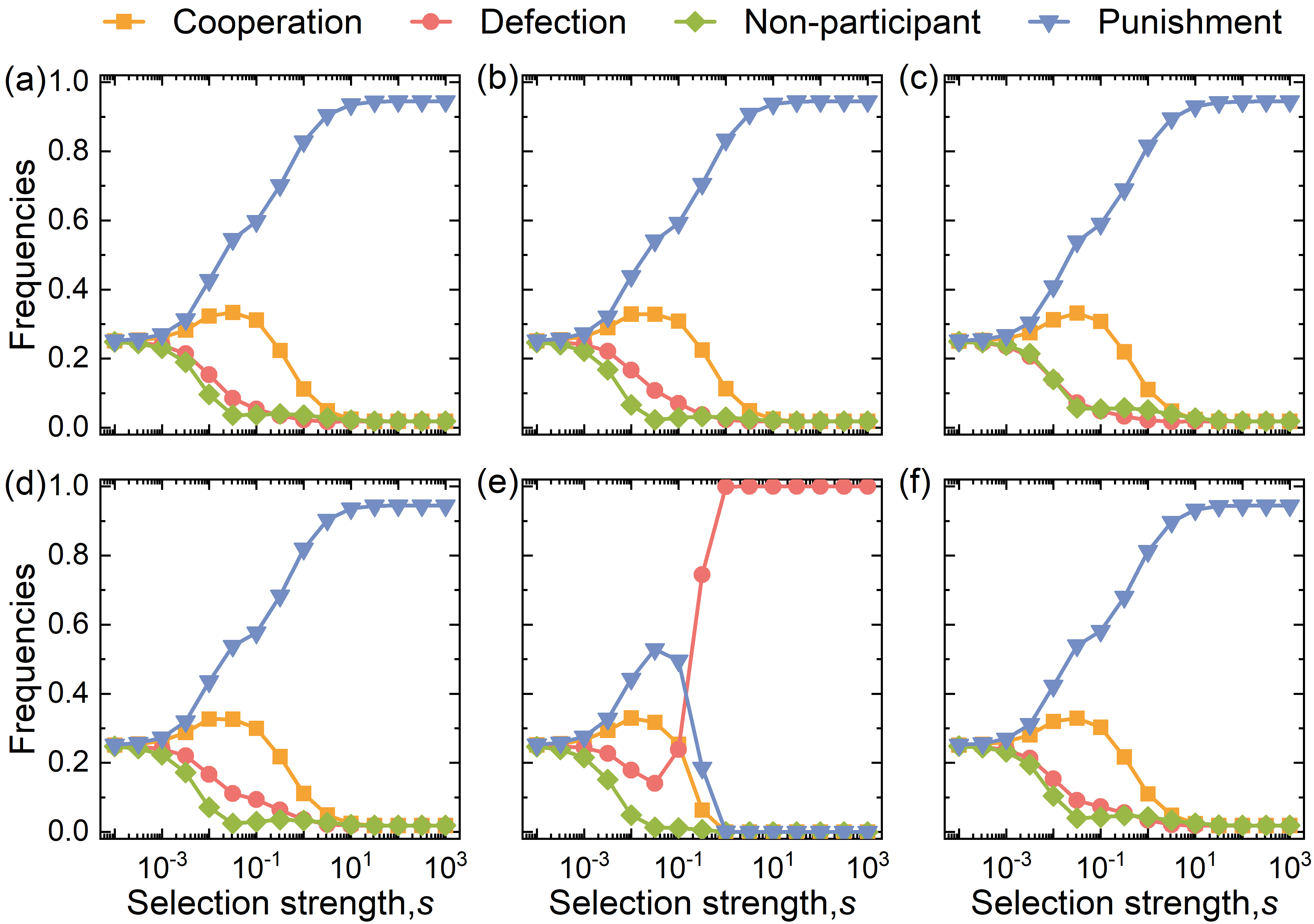}
    \caption{\textbf{The robustness across strong selection scenarios. }
    Shows are the stationary distribution as a function of selection strength $s$.
    Parameters are set as $r=4.5$, (a) $\sigma_1=1$ and $\sigma_2=0$, (b) $\sigma_1=1$ and $\sigma_2=1$, (c) $\sigma_1=1$ and $\sigma_2=-1$, (d) $\sigma_1=0.5$ and $\sigma_2=0$, (e) $\sigma_1=0.5$ and $\sigma_2=1$, and (f) $\sigma_1=0.5$ and $\sigma_2=-1$. }
    \label{fig:fig3_S2}
\end{figure}


\end{document}